\documentclass{article}

\usepackage{microtype}
\usepackage{graphicx}
\usepackage{subfigure}
\usepackage{booktabs,siunitx} 

\usepackage{hyperref}

\usepackage[accepted]{icml2023}

\usepackage{amsmath}
\usepackage{amssymb}
\usepackage{mathtools}
\usepackage{amsthm}

\usepackage[capitalize,noabbrev]{cleveref}

\theoremstyle{plain}
\newtheorem{theorem}{Theorem}[section]

\theoremstyle{definition}
\newtheorem{definition}[theorem]{Definition}

\theoremstyle{remark}

\usepackage[textsize=small]{todonotes}
\setuptodonotes{inline}

\usepackage{multirow}

\usepackage{makecell}
\usepackage{enumitem}
\usepackage{bm}
\usepackage{placeins}
\usepackage{pifont}
\newcommand{\cmark}{\ding{51}\xspace}%
\newcommand{\xmark}{\ding{55}\xspace}%

\usepackage{xspace}

\newcommand\name{PreNAS}
\newcommand\cnnname{PreNAS}
\newcommand\proxy{proxy}
\newcommand\proxies{proxies}
\newcommand\zerocost{zero-cost}

\icmltitlerunning{\name{}: Preferred One-Shot Learning Towards Efficient Neural Architecture Search}

\begin{document}

\twocolumn[
\icmltitle{\name{}: Preferred One-Shot Learning Towards \\
Efficient {\color{black}Neural Architecture Search}}



\icmlsetsymbol{equal}{*}

\begin{icmlauthorlist}
\icmlauthor{Haibin Wang}{equal,ali}
\icmlauthor{Ce Ge}{equal,ali}
\icmlauthor{Hesen Chen}{ali}
\icmlauthor{Xiuyu Sun}{ali}
\end{icmlauthorlist}

\icmlaffiliation{ali}{Alibaba Group, Beijing, China}

\icmlcorrespondingauthor{Xiuyu Sun}{xiuyu.sxy@alibaba-inc.com}

\icmlkeywords{Machine Learning, ICML}

\vskip 0.3in
]



\printAffiliationsAndNotice{\icmlEqualContribution} 

\begin{abstract}

The wide application of pre-trained models is driving the trend of once-for-all training in one-shot neural architecture search (NAS). However, training within a huge sample space damages the performance of individual subnets and requires much computation to search for an optimal model. In this paper, we present \name{}, a search-free NAS approach that accentuates target models in one-shot training. Specifically, the sample space is dramatically reduced in advance by a \zerocost{} selector, and weight-sharing one-shot training is performed on the preferred architectures to alleviate update conflicts. Extensive experiments have demonstrated that \name{} consistently outperforms state-of-the-art one-shot NAS competitors for both Vision Transformer and convolutional architectures, and importantly, enables instant specialization with zero search cost.  {\color{black}Our code is available at \href{https://github.com/tinyvision/PreNAS}{https://github.com/tinyvision/PreNAS}.}
 
\end{abstract}


\begin{figure}[t]
\vskip 0.1in
    \centering
    \includegraphics[width=0.95\linewidth]{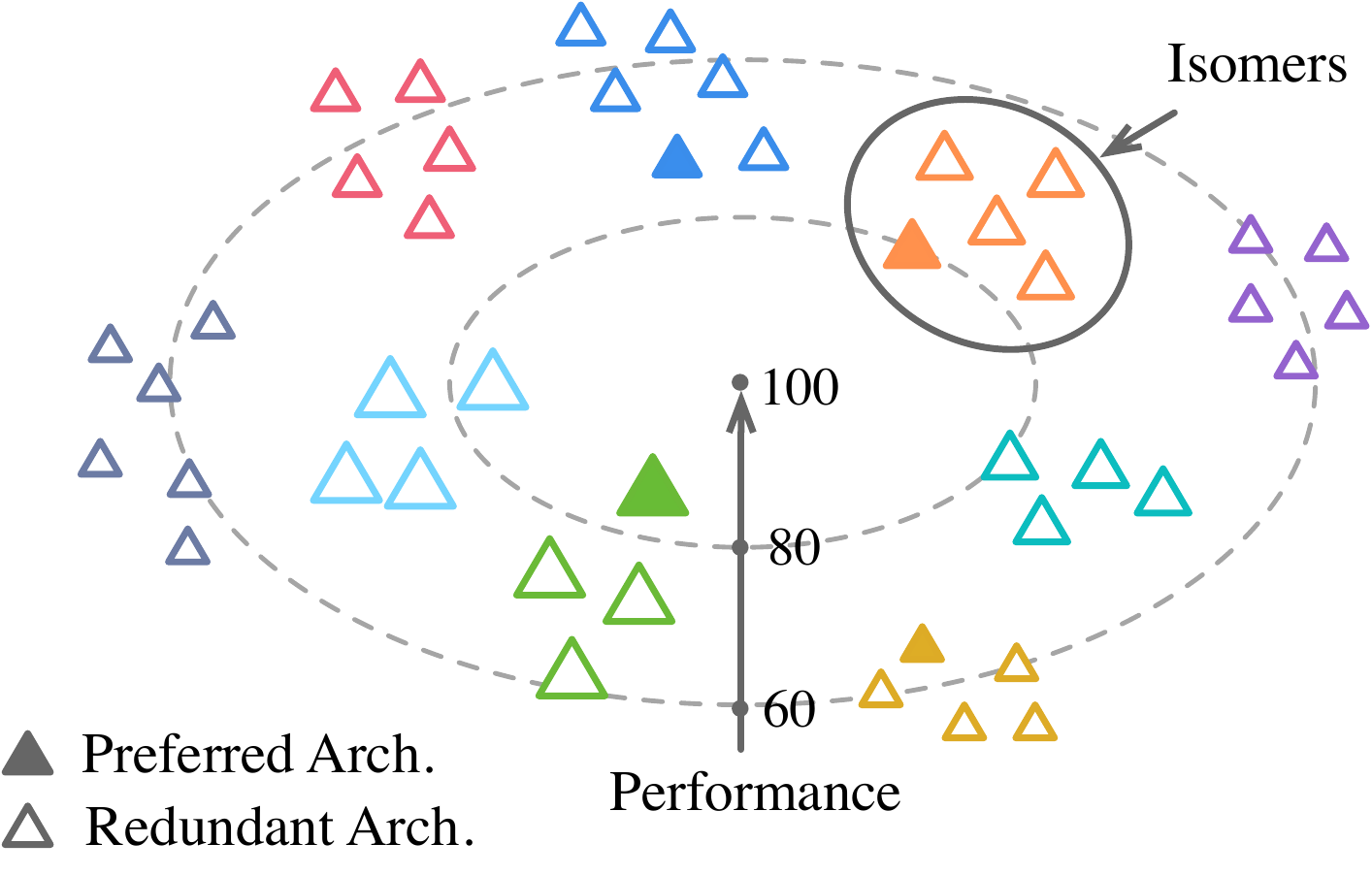}
    \caption{\name{} first identifies the preferred architectures via zero-cost \proxies{} within similar architectures, denoted as isomers, and then performs concentrated one-shot learning on only preferred architectures to achieve better convergence. The size of triangles implies different resource consumption of the architectures. }\label{fg:first_img}
\vskip -0.1in
\end{figure}

\begin{figure*}[t]
\vskip 0.1in
    \centering
    \includegraphics[width=0.9\textwidth]{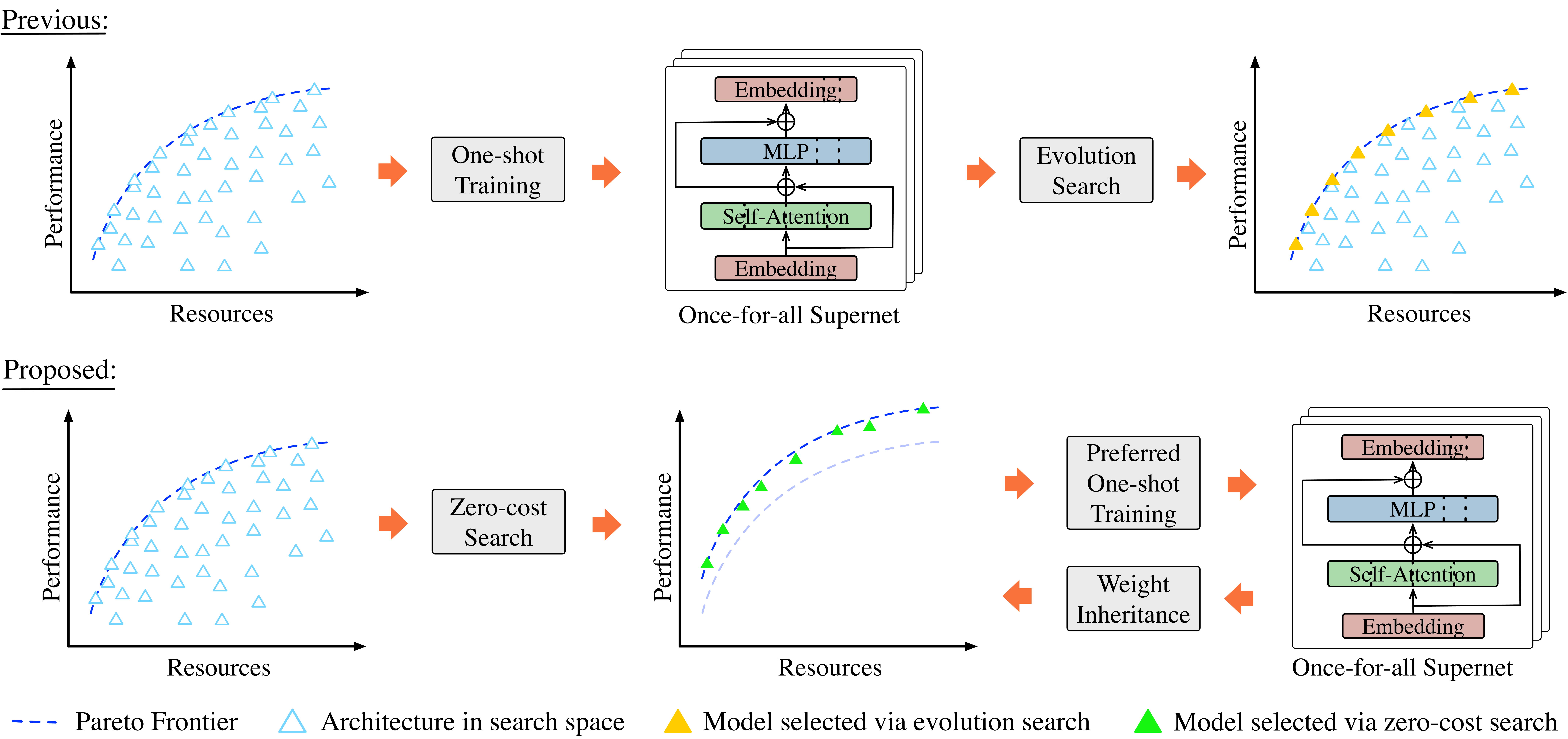}
    \caption{Illustration of \name{}. Previous one-shot NAS samples all architectures in the search space when one-shot training of the supernet for better evaluation in evolution search. Instead, \name{} first searches the target architectures via {\color{black}a} zero-cost proxy and next applies preferred one-shot training to supernet. \name{} improves the Pareto Frontier benefited from the preferred one-shot learning and is search-free after training by offering the models with the advance selected architectures from the zero-cost search.}
    \label{fg:overview}
\vskip -0.1in
\end{figure*}

\section{Introduction}
\label{introduction}

Deep learning has made significant strides in a wide range of tasks, including image classification~\cite{DBLP:conf/icml/TanL19, ViT}, speech recognition~\cite{hubert}, and natural language processing~\cite{bert}. One of the key factors contributing to their success is the design of model architectures, which can have a decisive impact on the final performance. However, manual design of model architectures is often time-consuming and requires significant expertise across different tasks. As a result, there is a growing need for automated design process. 

Neural architecture search (NAS) is a promising approach to address the challenges, with the goal of finding optimal architectures for a given task with minimal human intervention. One-shot NAS is a kind of NAS family that performs search through a well-trained weight-sharing model, known as a supernet. By training and evaluating possible subnets within a search space using only one single supernet, one-shot NAS significantly reduces the computational cost compared to other NAS methods that train and evaluate every model individually from scratch. One-shot NAS has demonstrated strong performance and has been applied to both convolutional neural networks (CNNs) and Transformer-based architectures~\cite{chitty2022neural}.

In order to reliably search for high-performance models, one-shot NAS solutions mostly employ a \textit{redundant} learning strategy. During the training phase, a plethora of subnets are sampled from a vast search space (can be $10^9$ or even much larger) to iteratively update the supernet; during the search phase, sufficient candidate subnets are populated to rank the optimal architecture. While this redundant strategy has indeed achieved good performance and reduced the uncertainty of NAS, it may also be limiting the cutting-edge breakthrough of one-shot NAS. To summarize, there are two specific issues to be concerned. (i) Search efficiency. Each search under the given resource constraints requires the evaluation of several thousand models, which is time-consuming and resource-intensive~\cite{cummings2022hardware}. (ii) Training efficacy. The vast training space makes learnable parameters within the supernet struggle to adapt to various subnets~\cite{focusformer}. The resultant gradient oscillations can impair training convergence and hurt the ultimate performance~\cite{gong2022nasvit}.

This work presents a novel \textit{preferred} learning strategy to fulfil the potential of one-shot NAS, namely \name{}. Considering practical application, NAS methods select high-performance models from the Pareto frontier under given resource constraints, as illustrated in Fig.~\ref{fg:first_img}. We are therefore inspired to pre-identify high-quality architectures with various resources in order to enable efficient access to optimal models, achieving nearly search-free specialization. Correspondingly, the selected candidates form a narrow sampling set, which we refer to as the preferred space. During training, the supernet is optimized on the preferred sample space to concentrate the training efficacy, which reduces gradient conflict and thus can potentially reach a better convergence state.

Identifying high-quality architectures without training is a non-trivial task. Another line of NAS research, zero-shot NAS, may provide a feasible solution in this regard. The core of zero-shot NAS lies in designing a \zerocost{} evaluation metric that serves as a fast proxy with high correlation to the actual performance. In this work, we creatively establish a linkage between one-shot NAS and zero-shot NAS to fully leverage the advantages of both. The key innovation is to enhance scoring-oriented proxy with design capability to handle isomeric Transformers. An undervalued but crucial phenomenon in NAS is that multiple architectures may own the same resource consumption, and this is particularly prevalent in Transformer architectures. The standard Vision Transformers are designed to have homogeneous embedding dimensions across blocks, which produces an interesting property that blocks can be rearranged to form new architectures while maintaining constant parameters and computation. Therefore, the evaluation proxy is revised to own the ability of eliminating isomeric architectures. Fig.~\ref{fg:first_img} provides a schematic diagram for the described process. Overall, the combination of one-shot and zero-shot NAS makes \name{} not only achieve rapid search of optimal neural architectures, but also obtain well-trained parameters efficiently, enabling out-of-the-box model specialization.

The main contributions are summarized as follows:
\begin{itemize}[nosep]
\item \name{}, a new learning paradigm that combines one-shot and zero-shot NAS, is proposed to train and search for optimal architectures within a preferred sample space, aiming to improve search efficiency and training efficacy.
\item A compound architecture selection mechanism is designed to construct and evaluate high-quality architectures, with a particular focus on addressing structural isomerism in Transformer architectures.
\item Extensive experiments and analysis are conducted to verify the effectiveness of \name{}, showing that it is competent to search for both CNN and ViT architectures with superior performance on various tasks.
\end{itemize}


\section{Background and Related Work}\label{background}

\subsection{One-Shot NAS}
In one-shot NAS, a weight-sharing supernet is built and trained to jointly optimize subnets within a pre-defined search space. Let $\mathcal{A}$ be the architecture search space and $W$ be all the learnable weights. The objective of one-shot training is to obtain the optimal weights
\begin{equation}\label{eq:originOneshotTraining}
W^* = \mathop{\arg \min}\limits_{W} \mathbb{E}^{}_{\alpha \sim U\{\mathcal{A}\}} [\mathcal{L}_\textup{train}(\alpha \mid W_\alpha, \mathcal{D}_\textup{train})],
\end{equation}
where subnet $\alpha$ is uniformly sampled from the search space $\mathcal{A}$, $W_\alpha$ is the corresponding part of inherited weights, and $\mathcal{L}_\textup{train}$ is a loss function defined on the training dataset. Upon the well-trained supernet, the best model $\alpha^*$ under the given resource constraint c is searched by ranking subnets as
\begin{equation}\label{eq:originEvolution}
\alpha^* = \mathop{\arg \max}\limits_{\alpha \in \mathcal{A}} \mathcal{P}_\textup{val}(\alpha \mid W^*_\alpha, \mathcal{D}_\textup{val}) \enspace \text{s.t.} \enspace r(\alpha) < c,
\end{equation}
where $\mathcal{P}_\textup{val}$ evaluates performance on the validation dataset and $r(\cdot)$ reports the resource consumption. During specialization, the search step can be conducted multiple times to obtain deploying models for different resource constraints.


The one-shot NAS~\cite{DBLP:conf/iclr/BrockLRW18} is proposed to amortize the cost from the training of each model at the beginning. It emphasizes high rank correlation of the accuracy with the quality of subnets~\cite{su2022vitas}, hence requires an additional retraining step after the best architecture is identified~\cite{wu2019fbnet,li2020block}. Recent works propose to alleviate this issue by training a once-for-all supernet in convolution networks~\cite{yu2019universally,DBLP:conf/iclr/CaiGWZH20,bignas}, Transformer~\cite{hat,autoformer} and hybrid networks of them~\cite{gong2022nasvit}. However, it is difficult to take account of both keeping the performance rank of all subnets in search space and achieving high performances of the target subnets in a once-for-all supernet. FocusFormer~\cite{focusformer} trains an architecture sampler instead of the uniform sampler in Eq.~\ref{eq:originOneshotTraining}, but does not fully address the problem. Instead, \name{} isolates the ranking task from the one-shot training via zero-shot NAS in advance.

\subsection{Zero-Shot NAS}
The advantage of one-shot NAS lies in its ability to simultaneously return both the architectures and optimized parameters. It is more suitable for scenarios where pre-trained models are required, e.g., AutoTinyBERT~\cite{autotinybert} for BERT~\cite{bert} and LightHuBERT~\cite{lighthubert} for HuBERT~\cite{hubert}. However, when only architectures are needed, one-shot NAS would be over cumbersome.

Zero-shot NAS seeks to find high-quality architectures without costly model training. Various \zerocost{} proxies have been proposed to efficiently estimate architecture quality. The general purpose can be formulized as
\begin{equation}\label{eq:zeroCost}
\alpha^* = \mathop{\arg \max}\limits_{\alpha \in \mathcal{A}} \mathcal{P}_\textup{score}(\alpha) \quad \text{s.t.} \quad r(\alpha) < c ,
\end{equation}
where $\mathcal{P}_\textup{score}$ can rapidly score architectures without training and validation on large-scale datasets.


For instance, \citet{DBLP:conf/icml/MellorTSC21} have developed a zero-cost \proxy{} by the overlap of activations between datapoints in untrained networks. \citet{zennas} propose Zen-Score to represent the network expressivity. Meanwhile, previous works have leveraged the high correlation of the gradient-based zero-cost \proxies{} with the model performances, such as the gradient norm~\cite{DBLP:conf/iclr/AbdelfattahMDL21}, SNIP~\cite{snip}, GraSP~\cite{grasp}, SynFlow~\cite{synflow}, NASI~\cite{nasi}. Further, \citet{shu2022unifying} theoretically prove the connections among different gradient-based zero-cost \proxies{} and propose HNAS to consistently boost existing training-free NAS algorithms. \citet{knas} apply their proposed gradient-based \proxy{} to RoBERTa~\cite{liu2019roberta}. TF-TAS~\cite{tftas} is the first gradient-based zero-cost \proxy{} especially for ViT~\cite{ViT}. \citet{javaheripilitetransformersearch} surprisingly find that the number of decoder parameters in autoregressive Transformers has a high rank correlation with task performance.

Without loss of generality, SNIP~\cite{snip}, a representative gradient-based proxy, is adopted in this work. The scoring function of SNIP is defined as:
\begin{equation}\label{eq:snip}
\mathcal{P}_\textup{\tiny SNIP}(\alpha) = \sum_{\theta_i \in \theta_\alpha} \left| \frac{\partial\mathcal{L}}{\partial\theta_i}\odot\theta_i \right|.
\end{equation}
We use $\theta$ to denote randomly initialized parameters to be differentiated from trained weights $W$. In order to define the loss function $\mathcal{L}$, at least one mini-batch data is required to perform a forward inference. SNIP can be freely replaced with other \zerocost{} proxies as long as they are competent to select high-quality architectures. As shown in Fig.~\ref{fg:overview}, the models sampled by zero-cost proxy are typically good enough with only a small deviation from the Pareto Frontier.



\section{\name{}}\label{method}

In this section, we begin by revealing the underlying potential of conventional one-shot NAS, and introduce a new preferred learning strategy to fully release its capability as Fig.~\ref{fg:overview} shows. Then we discuss the isomeric issue in Transformer architectures and detail the adapted zero-cost proxy. Finally, we propose addressing training fairness to balance performance distribution.

\subsection{One-Shot NAS with Preferred Learning}

As mentioned earlier, conventional one-shot NAS has shortcomings in terms of search efficiency and training efficacy. During the training phase, Eq.~\ref{eq:originOneshotTraining} expects every subnet $\alpha \in \mathcal{A}$ to be sampled and optimized to reach matchable performance with the architecture quality. However, the search space $\mathcal{A}$ is almost infinitely large, and even redundant training strategy cannot traverse all subnet architectures. More importantly, since the learnable weights $W$ are shared across high-quality and low-quality subnets, the final convergence state is weakened due to conflicting adaptations. Besides, to search for the best architecture from a group of candidate architectures, the performance evaluator $\mathcal{P}_\textup{val}$ in Eq.~\ref{eq:originEvolution} could be executed several thousand times in a single search. This search-time overhead is inevitable even with improved strategies like random search~\cite{DBLP:conf/icml/BenderKZVL18, DBLP:conf/uai/LiT19}, evolution algorithm~\cite{DBLP:conf/aaai/RealAHL19, DBLP:conf/eccv/GuoZMHLWS20} and reinforcement learning~\cite{pham2018efficient, DBLP:conf/cvpr/TanCPVSHL19}.


From the perspective of improving search efficiency, the number of candidates to be evaluated should be minimized. A beneficial side effect of this treatment is that it allows for focused updates on fewer subnets, and potentially enables each model to reach a better convergence state. Therefore, we propose performing one-shot learning within a much smaller preferred search space $\widetilde{\mathcal{A}}$ that which satisfies
\begin{equation}
    \widetilde{\mathcal{A}} \subset \mathcal{A} \enspace \text{and} \enspace |\widetilde{\mathcal{A}}| \ll |\mathcal{A}| .
\end{equation}
The reduced search space should be of sufficient quality, so that the best architecture found under the given resource constraint is comparable to the theoretically optimal one in the entire search space, i.e., 
\begin{align}
    \max\limits_{\tilde{\alpha} \in \mathcal{\Tilde{A}}} \mathcal{P}_\textup{val}(\tilde{\alpha}) & \approx  \max\limits_{\alpha \in \mathcal{A}} \mathcal{P}_\textup{val}(\alpha)
    \shortintertext{subject to}
    \quad r(\tilde{\alpha}) < c \enspace & \text{and} \enspace r(\alpha) < c.
\end{align}
The symbols of weight and dataset in $\mathcal{P}_\textup{val}$ are omitted for simplicity. 

More clearly, the preferred search space is formed by selecting high-quality architectures under various resource constraints:
\begin{equation}\label{eq:selector}
    \widetilde{\mathcal{A}} = \left\{\, \mathcal{S}(\mathcal{A}_{[c]}, N) \mid \forall c \in \mathcal{C} \right\}.
\end{equation}
The selector $\mathcal{S}$ chooses the top-N best architectures from the valid subsets and $\mathcal{A}_{[c]}$ is defined as 
\begin{equation}
    \mathcal{A}_{[c]} = \left\{\, \alpha \mid \alpha \in \mathcal{A} \enspace \text{s.t.} \enspace r(\alpha) < c \right\}.
\end{equation}
Constraints $\mathcal{C}$ are a series of limitations on resource consumption, e.g., model size or FLOPs,
\begin{equation}
\mathcal{C} = (a, a+\varepsilon, a+2\varepsilon, \cdots, b),
\end{equation}
where $\varepsilon$ is the discretization margin and $a, b$ are the lower and upper bounds of the search space. 


Upon the preferred search space, the supernet is trained in a weight-entanglement manner~\cite{autoformer} but focusing on only high-quality architectures. The preferred one-shot training process can be rewritten as:
\begin{equation}\label{eq:our_train}
\widetilde{W}^* = \mathop{\arg \min}\limits_{\widetilde{W}} \mathbb{E}^{}_{\alpha \sim B\{\widetilde{\mathcal{A}}\}} [\mathcal{L}_\textup{train}(\alpha \mid \widetilde{W}_\alpha, \mathcal{D}_\textup{train})].
\end{equation}
Concentrated optimization on high-quality subnets avoids sharing weights with poor architectures and results in superior individual performance, which is confirmed by our state-of-the-art results reported in Section~\ref{sec:sota}.

In the specialization stage, the optimal architectures under given  resource constraints can be directly obtained:
\begin{equation}
\widetilde{\mathcal{A}}^* = \widetilde{\mathcal{A}} .
\end{equation}
This procedure does not require costly search and evaluation on a large number of candidates. It achieves zero-cost specialization in most cases and is sufficient for general uses. For rare finer-grained use cases, there is only a small additional overhead to rank a few subnets.



\subsection{Zero-Cost Transformer Selector}\label{sub:layerNorm}



\paragraph{Proxy Confusion} The foundation of \name{} lies in selecting high-quality architectures before data-based training, i.e., the selector $\mathcal{S}$ in Eq.~\ref{eq:selector}. This is a non-trivial task and we resort to the proxy techniques in zero-shot NAS research. A naive idea would be to use gradient-based proxies such as SNIP~\cite{snip} to search for qualified subnets. After preliminary experiments, we found that SNIP as well as other similar proxies perform well across architectures of varying capacity. The computed scores have a high correlation to true accuracy. However, it encounters confusion in distinguishing architectures with the same resource consumption. 


\begin{definition}[Transformer isomers]
For a $L$-block Transformer architecture $\alpha \coloneqq (\beta_1, \beta_2, \dots, \beta_L)$ with $\beta_i$ the configurations of $i$-th block, all architectures produced by reordering $\alpha$ are isomers.
\end{definition}
 
\begin{figure}[t]
 \vskip 0.1in
    \centering
    \subfigure[SNIP]{\includegraphics[width=0.49\linewidth]{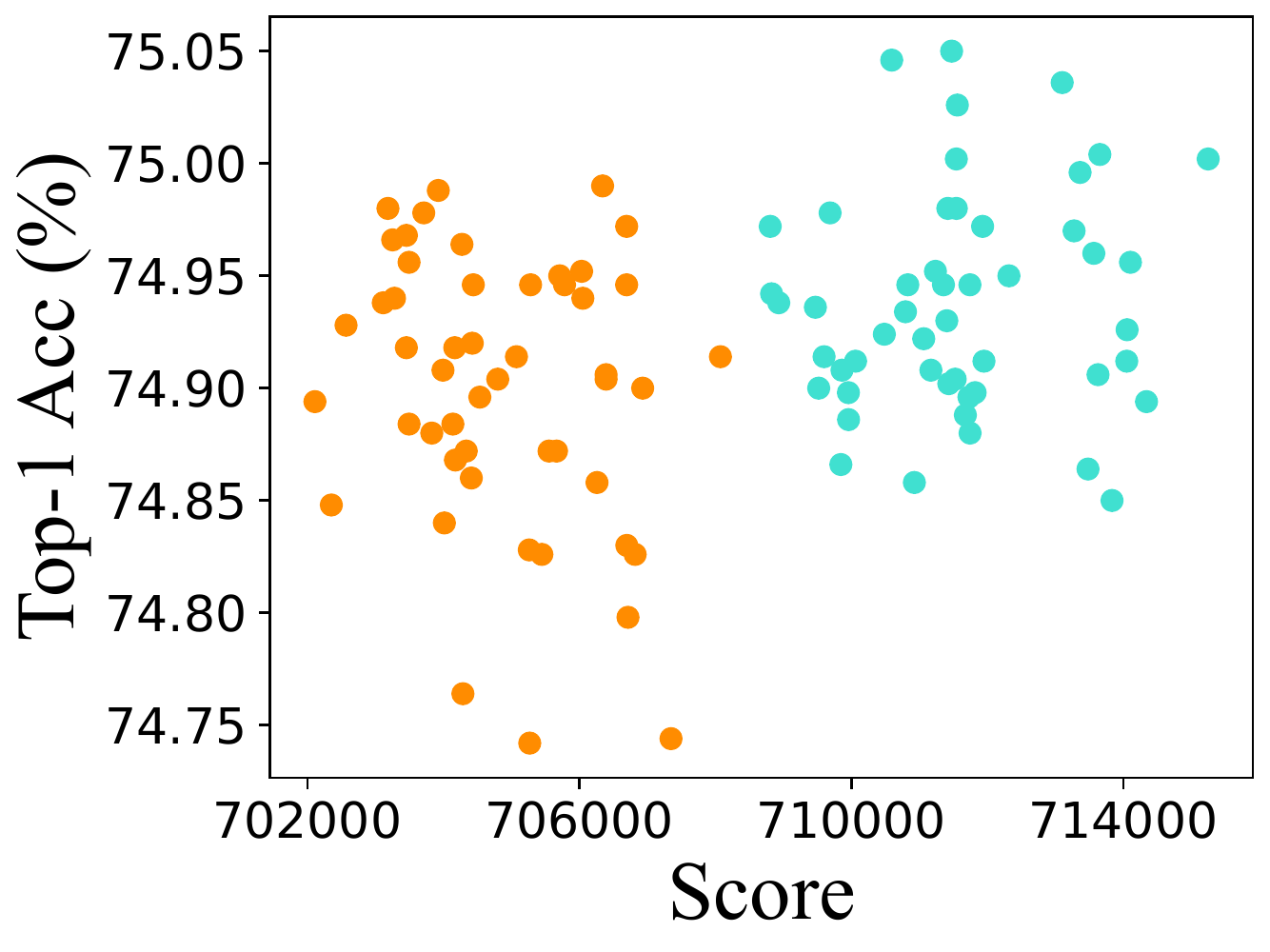}\label{fg:correlation:a}}%
    \hfil%
    \subfigure[Normalized SNIP]{\includegraphics[width=0.49\linewidth]{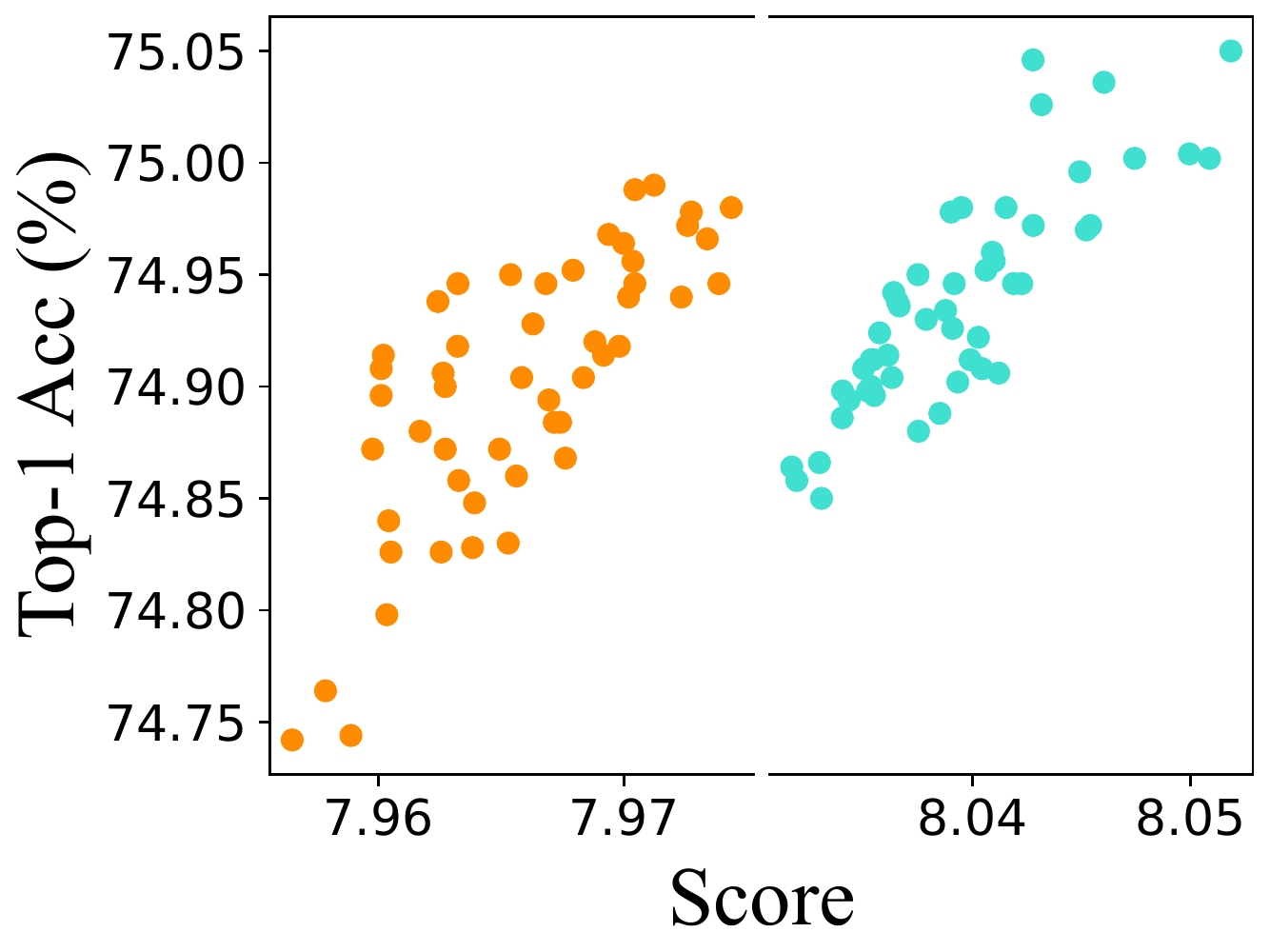}\label{fg:correlation:b}}
    \subfigure[Score sensitivity]{\includegraphics[width=0.98\linewidth]{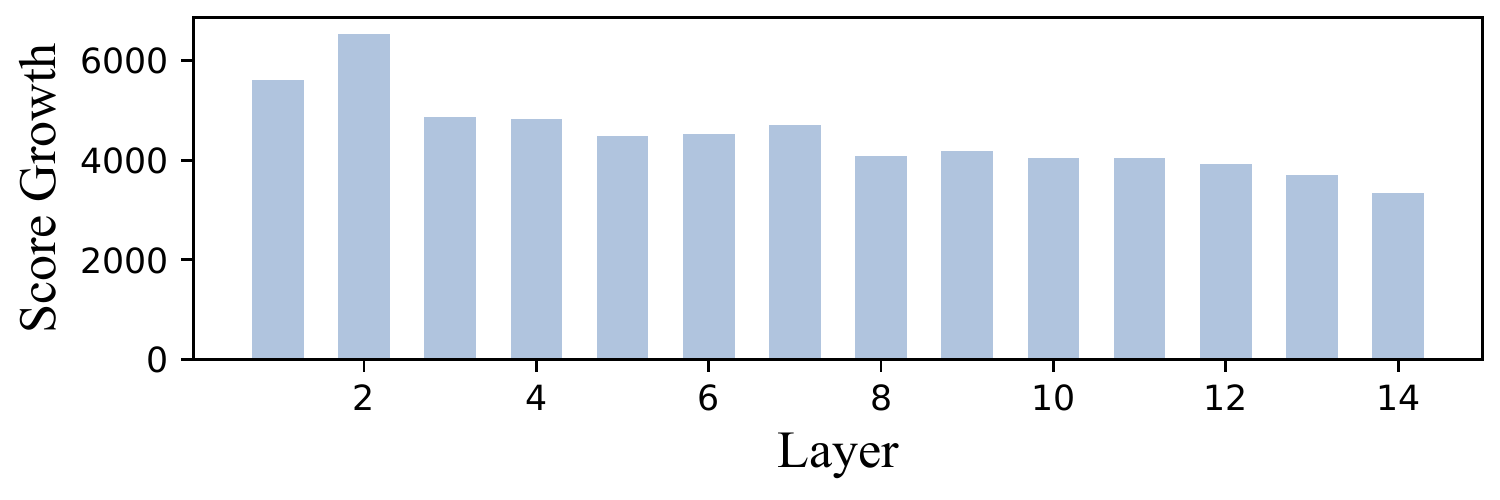}\label{fg:correlation:c}}
    \caption{(a) The SNIP scores are uncorrelated with the test accuracy of various subnets in two isomers. (b) The SNIP scores with the layer normalization have significant positively correlation with the test accuracy of various subnets in the two isomers. (c) The growth of the SNIP score in each layer with the increase of the MLP ratio from 3.5 to 4.}\label{fg:correlation}%
\vskip -0.1in
\end{figure}

\paragraph{Understanding Isomers} We attribute the confusion phenomenon to the isomerism of Transformer architectures. SNIP was originally proposed for CNN architectures and shows a high correlation with model size. However, Transformer isomers have constant model sizes and FLOPs, which deceives the SNIP score. Fig.~\ref{fg:correlation:a} illustrates two groups of isomers, and it is clear that there is no significant correlation between SNIP and accuracy. By inspecting per-block scores of the supernet, we observe from Fig.~\ref{fg:correlation:c} that the unit contribution of lower blocks is relatively higher. Therefore, SNIP tends to select architectures with wider lower blocks, which typically exhibit mediocre performance. Refer to Appendix~\ref{appendix:snipAnalysis} for more detailed analyses. It is noteworthy that the slight advantage of lower blocks can be easily dominated by scaling embedding width and the number of blocks, which explains why the SNIP still works under different resource consumption.


\paragraph{Normalized Isomer Proxy} Based on the above analysis, we propose a natural solution that normalizes SNIP on a per-layer basis. The revised scoring function of normalized SNIP is defined as:
\begin{equation}\label{eq:snipLayernorm}
\mathcal{P}\strut^\textup{\tiny norm}_\textup{\tiny SNIP} = \sum^L_{l=1}\left( \sum_{\theta_\alpha^l}\left|\frac{\partial\mathcal{L}}{\partial\theta_\alpha^l}\odot\theta_\alpha^l\right|  \bigg/  \sum_{\theta^l}\left|\frac{\partial\mathcal{L}}{\partial\theta^l}\odot\theta^l\right| \right),
\end{equation}
where $\theta^l$ denotes the randomly initialized parameters of the supernet at $l$-th block and $\theta_\alpha^l$ is the inherited parameters of subnet $\alpha$. As Fig.~\ref{fg:correlation:b} shows, the normalized SNIP score shows a positive correlation with accuracy for isomers. It is noteworthy that each isomer group may contain a large number of architectures, and applying $\mathcal{P}^\textup{\tiny norm}_\textup{\tiny SNIP}$ on all of them is inefficient. Other approximation techniques like random search or evolutionary algorithm can help speed up but do not guarantee always getting the best results. Therefore, we provide an efficient allocation algorithm based on $\mathcal{P}^\textup{\tiny norm}_\textup{\tiny SNIP}$ that greedily constructs the top-scoring isomer in $\mathcal{O}(1)$ complexity. The algorithm is detailed in Appendix~\ref{appendix:greedy}.

Finally, the zero-cost selector $\mathcal{S}$ in Eq.~\ref{eq:selector} is a compound two-step operation. It first apply $\mathcal{P}^\textup{\tiny norm}_\textup{\tiny SNIP}$ on isomers to elect representatives and then apply $\mathcal{P}^{}_\textup{\tiny SNIP}$ to select the top-$N$ preferred architectures under the given resource constraint.

\subsection{Performance Balancing}

We discuss training fairness and demonstrate how to distribute learning performance more fairly across subnets. As indicated in Eq.~\ref{eq:originOneshotTraining}, the usual one-shot training  samples subnets uniformly from the vast search space, i.e., $\alpha \sim U\{\mathcal{A}\}$. In our preferred search space $\widetilde{\mathcal{A}}$, the candidate architectures are evenly distributed in terms of resource consumption. However, a uniform sampling on $\widetilde{\mathcal{A}}$ do not result in fair updates in terms of subnets.  The variable building factors like embedding dimension, number of blocks, and number of attention heads do not appear equally, among which the former two have a greater impact. Hence, we propose grouping the architectures in $\widetilde{\mathcal{A}}$ by both their embedding dimensions and depths and perform uniform sampling on this calibrated distribution, i.e., $\alpha \sim B\{\widetilde{\mathcal{A}}\}$ in Eq.~\ref{eq:our_train}.


\section{Experiments}\label{experiments}


\begin{table}[t]
\caption{The search space of \name{} for Vision Transformer. Following AutoFormer~\cite{autoformer}, three supernets are built spanning different parameter scales. The triplet of each variable factor means the lower bound, upper bound, and incremental step, respectively. The Q-K-V dimensions, numbers of heads, and MLP ratios varies across different layers.
}\label{tab:AutoFormerSearchSpace}
\vskip 0.1in
\begin{center}
\begin{footnotesize}
\setlength{\tabcolsep}{0.2em}
\begin{tabular*}{\linewidth}{l| @{\extracolsep{\fill}} ccc}
\toprule
& Supernet-Tiny  &  Supernet-Small  & Supernet-Base \\
\midrule
Embed Dim & (192, 240, 24) & (320, 448, 64) & (528, 624, 48) \\
Q-K-V Dim & (192, 256, 64) & (320, 448, 64) & (512, 640, 64) \\
MLP Ratio & (3.5, 4, 0.5)  & (3, 4, 0.5) & (3, 4, 0.5) \\
Head Num & (3, 4, 1) & (5, 7, 1) & (8, 10, 1) \\
Depth Num & (12, 14, 1) & (12, 14, 1) & (14, 16, 1) \\
\midrule
Params Range & 5.4 -- 10.5M & 13 -- 34M & 42 -- 76M \\
\bottomrule
\end{tabular*}
\end{footnotesize}
\end{center}
\vskip -0.1in
\end{table}



\subsection{Setup}

\paragraph{Search Space}

We employ the same search space of AutoFormer~\cite{autoformer} including five variable factors in Transformer blocks: embedding dimension, Q-K-V dimension, number of heads, MLP ratio, and network depth, as shown in Tab.~\ref{tab:AutoFormerSearchSpace}. We also apply the paradigm of \name{} to CNNs with the same search space of BigNAS~\cite{bignas} in Tab.~\ref{tab:BigNASSearchSpace}.


\paragraph{Implementation Details} 

We implemented \name{} upon the PyTorch~\cite{pytorch} framework with improvements from the timm~\cite{timm} library. The supernets are trained following the recipe outlined in AutoFormer~\cite{autoformer}, where multiple data augmentation and regularization techniques are utilized to facilitate convergence, including RandAugment~\cite{randaug}, mixup~\cite{mixup}, CutMix~\cite{cutmix}, Random Erasing~\cite{random_erasing}, stochastic depth~\cite{droppath}, Repeated Augmentation~\cite{multigrain,batchaug} and Label Smoothing~\cite{label_smoothing_2016,label_smoothing_2020}. The detailed hyper-parameters are presented in Appendix~\ref{appendix:aug}. The input images are all resized to $224\times244$ and split into patches of size $16\times16$. We use the AdamW optimizer with a mini-batch size of 1024. The learning rate is initially set to 1e-3 and decays to 2e-5 through a cosine scheduler in 500 epoches. The discretization margin $\varepsilon$ is set to 1M. We conducted experiments and measured design time on NVIDIA A100 GPUs.

\begin{table}[t]
\caption{The search space of \name{} based on MobileNetV2. Following BigNAS~\cite{bignas}, the building blocks include vanilla convolutions and inverted bottleneck residual blocks (MBConv)~\cite{howard2018inverted}.
}\label{tab:BigNASSearchSpace}
\vskip 0.1in
\begin{center}
\begin{footnotesize}
\setlength{\tabcolsep}{0.2em}
\begin{tabular*}{\linewidth}{c| @{\extracolsep{\fill}} ccccc}
\toprule
Stage & Operator & Resolution & \#Channels & \#Layers & Kernel \\
\midrule
& Conv & 192 -- 320 & 32 -- 40 & 1 & 3 \\
1 & MBConv1 & 96 -- 160 & 16 -- 24 & 1 -- 2 & 3 \\
2 & MBCouv6 & 96 -- 160 & 24 -- 32 & 2 -- 3 & 3 \\
3 & MBCouv6 & 48 -- 80 & 40 -- 48 & 2 -- 3 & 3, 5 \\
4 & MBCouv6 & 24 -- 40 & 80 -- 88 & 2 -- 4 & 3, 5 \\
5 & MBCouv6 & 12 -- 20 & 112 -- 128 & 2 -- 6 & 3, 5 \\
6 & MBCouv6 & 12 -- 20 & 192 -- 216 & 2 -- 6 & 3, 5 \\
7 & MBCouv6 & 6 -- 10 & 320 -- 352 & 1 -- 2 & 3, 5 \\
& Conv & 6 -- 10 & 1280 -- 1408 & 1 & 1 \\
\bottomrule
\end{tabular*}
\end{footnotesize}
\end{center}
\vskip -0.1in
\end{table}

\subsection{Main Results}\label{sec:sota}

\paragraph{Comparison with NAS ViT} The results of \name{} are presented in Tab.~\ref{tab:sota}. In accordance with DeiT~\cite{deit}, we compare the best results on three specifications, viz., \name{}-Tiny, \name{}-Small, and \name{}-Base, each corresponding to a parameter limit of 6M, 23M, and 54M, respectively. The recent AutoFormer~\cite{autoformer} is the most relevant NAS competitor, which adopted a redundant one-shot training strategy and searched for optimal subnets by evolutionary algorithm. It can be observed that our \name{} surpasses AutoFormer on both top-1 and top-5 accuracy under all three resource constraints. Besides, it is important that these achievements are obtained with highly competitive search efficiency. Specifically, AutoFormer utilizes evolutionary algorithm and requires at least 90 GPU hours to safely provide an optimal model. While \name{} is indeed search-free after one-shot training and can instantly provide a desired model under the given resource constraint.

\paragraph{Comparison with Handcrafted ViT} Some competitive ViT models that were elaborately designed by humans are also included, among which DeiT~\cite{deit}, ConViT~\cite{convit}, TNT~\cite{tnt} and T2T-ViT~\citet{t2t} are pure ViT architectures, while LVT~\cite{lvt} and BoTNet~\cite{botnet} are hybrid architectures built of convolutions and Transformer blocks. While boosting accuracy, \name{} also tend to have fewer parameters and FLOPs than manual designs, especially in larger-scale models. For example, \name{}-Base improves top-1 accuracy from 81.8\% to 82.6\% with only 63\% parameters and 61\% FLOPs of DeiT-B. It is noteworthy that hybrid architectures utilize convolutions to effectively down-sample features, which yields significant advantageous in computation reduction. Surprisingly, \name{} demonstrates comparable performance to BoTNet with similar or even fewer parameters and FLOPs. 

In summary, the experimental results firmly reveal the effectiveness of \name{}. It not only outperforms pure state-of-the-art Vision Transformers, but also reaches an extraordinary level of parity with compact hybrid architectures.
\begin{table}[tb]
\caption{Comparison of different Vision Transformers on ImageNet. ${}^\ddagger$Hybrid models of convolutions and Transformer blocks.}\label{tab:sota}
\vskip 0.1in
\begin{center}
\begin{small}
\begin{sc}
\setlength{\tabcolsep}{0.2em}
\begin{tabular*}{\linewidth}{l@{\extracolsep{\fill}}ccccc}
\toprule
Model                && \makecell{Top-1\\ (\%)}        & \makecell{Top-5\\ (\%)}       & \makecell{\#Params\\ (M)}      & \makecell{FLOPs\\ (G)}        \\ \midrule
LVT$^\ddagger$                   && 74.8          & 92.6          & 5.5           & 0.9          \\
DeiT-Ti               && 72.2          & 91.1          & 5.7           & 1.2          \\
ConViT-Ti             && 73.1          & 91.7          & 6.0           & 1.0          \\
TNT-Ti                && 73.9          & 91.9          & 6.1           & 1.4          \\
AutoFormer-Ti       && 74.7          & 92.6          & 5.9           & 1.3          \\
\textbf{\name{}-Ti}     && \textbf{77.1} & \textbf{93.4} & \textbf{5.9}  & \textbf{1.4} \\
\midrule
BoTNet-S1-59$^\ddagger$          && 81.7          & 95.8          & 33.5          & 7.3          \\
DeiT-S                && 79.8          & 95.0          & 22.1          & 4.7          \\
ConViT-S              && 81.3          & 95.7          & 27.0          & 5.4          \\
TNT-S                 && 81.5          & 95.7          & 23.8          & 5.2          \\
T2T-ViT-14            && 81.7          & -             & 21.5          & 6.1          \\
AutoFormer-S          && 81.7          & 95.7          & 22.9          & 4.9          \\
\textbf{\name{}-S}     && \textbf{81.8} & \textbf{95.9} & \textbf{22.9} & \textbf{5.1} \\
\midrule
BoTNet-S1-110$^\ddagger$         && 82.8          & 96.3          & 55            & 11           \\
DeiT-B                && 81.8          & 95.6          & 86            & 18           \\
ConViT-B              && 82.4          & 95.9          & 86            & 17           \\ 
AutoFormer-B          && 82.4          & 95.7          & 54            & 11           \\
\textbf{\name{}-B}     && \textbf{82.6}          & \textbf{96.0}          & \textbf{54}   & \textbf{11}    \\
\bottomrule
\end{tabular*}
\end{sc}
\end{small}
\end{center}
\vskip -0.1in
\end{table}

\begin{table}[t]
\caption{Performance comparison of subnets trained from scratch (300 epochs and 1024 batch size). SNIP is applied by replacing the normalized SNIP in the first step of seletor $\mathcal{S}$. The subscript $_{ZS}$ means using only the zero-shot selector of \name{}.}\label{tab:zeroshot}
\vskip 0.1in
\begin{center}
\begin{small}
\begin{sc}
\setlength{\tabcolsep}{0.2em}
\begin{tabular*}{\linewidth}{l@{\extracolsep{\fill}}ccccc}
\toprule
Method                && \makecell{Top-1\\ (\%)}      & \makecell{Top-5\\ (\%)}    & \makecell{\#Params\\ (M)} & \makecell{Time\\ (hours)}   \\ \midrule
SNIP-Ti           && 74.4          & 92.4     & 6.0 &  0.1   \\
AutoFormer-Ti        && 74.4          & 92.4       & 5.9 & 415  \\
TF-TAS-Ti               && 75.2          & 92.7     & 6.2 & 12 \\
\textbf{\name{}-Ti} $_{ZS}$        && \textbf{74.8}          & \textbf{92.6}    & \textbf{6.0} & \textbf{0.1} \\ \midrule
SNIP-S          && 81.2          & 95.7    & 23 &  0.3  \\
AutoFormer-S     && 81.5          & 95.6     &  23 & 667   \\
TF-TAS-S               && 81.4           & 95.7    & 24 & 17  \\
\textbf{\name{}-S} $_{ZS}$        && \textbf{81.5}          & \textbf{95.8}   & \textbf{23} & \textbf{0.3}  \\ \bottomrule
\end{tabular*}
\end{sc}
\end{small}
\end{center}
\vskip -0.1in
\end{table}

\subsection{Analysis and Ablation study}\label{ablation}

\paragraph{Quality of Preferred Architectures} In \name{}, the quality of preferred architectures $\widetilde{\mathcal{A}}$ is crucial and is the basis of the entire framework. Therefore, we take the architectures selected by zero-cost selector $\mathcal{S}$ (without subsequent one-shot training) and train them from scratch to solely verify its effectiveness. The results are shown in Tab.~\ref{tab:zeroshot} and all the methods are fairly trained for 300 epochs with the same recipe. In both the Tiny and Small search spaces, the trained accuracy of \name{} is superior to other competitors, except for TF-TAS-Ti, which actually has advantageous more parameters. {\color{black} We attribute the benefits of PreNAS to its normalization of isomer proxy, which eliminates the preference for wider lower blocks in SNIP. For example, the MLP ratios from low level layer to high level layer in SNIP-Ti are 4.0, 4.0, 4.0, 4.0, 4.0, 4.0, 4.0, 4.0, 4.0, 3.5, 3.5, 3.5 respectively, while they are 4.0, 4.0, 4.0, 4.0, 4.0, 3.5, 3.5, 4.0, 4.0, 3.5, 4.0, 4.0 in PreNAS-Ti.} From the perspective of efficiency, our two-step selector is competent as an excellent zero-shot proxy. The search time is nearly zero at the same level as SNIP, and the architecture quality is comparable or even better than evolutionary search in one-shot AutoFormer. TF-TAS~\cite{tftas} is a recently proposed zero-shot proxy specifically for Transformers. Although it promoted the search efficiency by $48\times$ compared to AutoFormer, it still consumes at least 12 hours to offer an architecture via massive forward inferences. In contrast, \name{} requires only one forward and backward propagation via a mini batch and all subnets directly inherit the initialized weights and gradients from supernet.

\begin{figure}[t]
\vskip 0.1in
    \centering
    \begin{minipage}[t]{0.48\linewidth}
    \includegraphics[width=\linewidth]{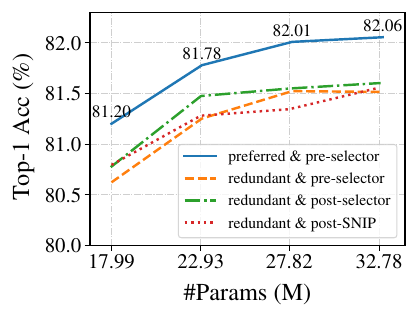}%
    \caption{ImageNet accuracy of different routines of zero-cost proxies and one-shot training in the Small space.}\label{fg:searchTrain}
    \end{minipage}%
    \hfill%
    \begin{minipage}[t]{0.48\linewidth}
    \includegraphics[width=\linewidth]{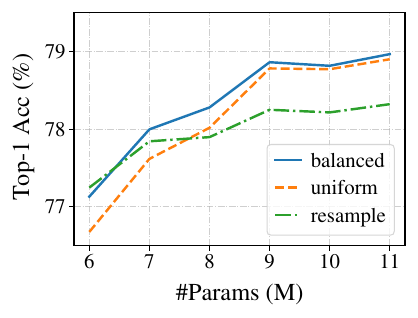}%
    \caption{ImageNet accuracy of different sampling strategies during one-shot training in the Tiny space.}\label{fg:balancing}
    \end{minipage}%
\vskip -0.1in
\end{figure}

\paragraph{Investigation of Preferred Training} \name{} is a combination of zero-shot selector and preferred one-shot training. It is possible to combine the proposed selector with conventional redundant one-shot training. Fig.~\ref{fg:searchTrain} shows several possible alternatives. ``preferred $\&$ pre-selector'' is the recipe of \name{}. ``redundant $\&$ pre-selector'' means training supernet in the redundant manner and selecting subnets beforehand with randomly initialized parameters, while ``redundant $\&$ post-selector'' means that subnets are selected afterwards from supernet with well-trained parameters. The figure confirms the effectiveness of preferred training in convergence. Specifically, with the same selected subnets in ``preferred $\&$ pre-selector'' and ``redundant $\&$ pre-selector'', the preferred training improves accuracy by about 0.5\% for each subnet comparing to the redundant ones. By concentrating on high-quality architectures, the preferred training encounters less update conflicts and thus achieves more adequate optimization for all selected subnets.

\begin{figure}[t]
\vskip 0.1in
    \centering
    \subfigure[Search spaces]{\includegraphics[width=0.36\linewidth]{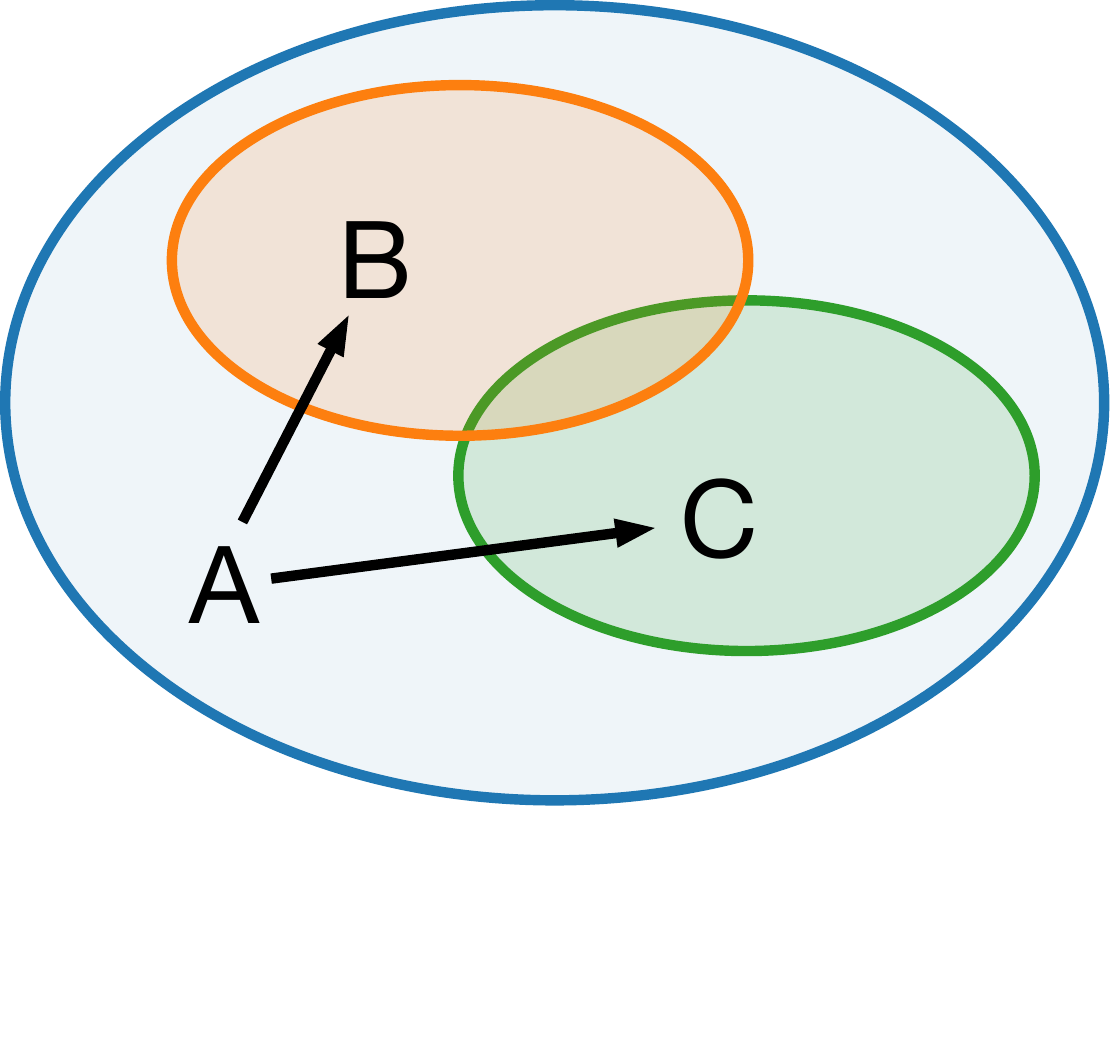}\label{fg:search_space:a}}%
    \hfil%
    \subfigure[Error distributions]{\includegraphics[width=0.58\linewidth]{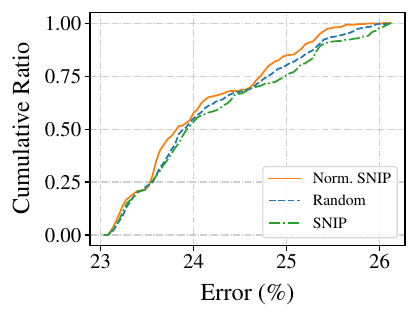}\label{fg:search_space:b}}
    \caption{Model quality in different search spaces. (a) Space $A$ is the initial entire space, $B$ is the refined search space by applying layer-normalized SNIP to isomers as in Eq.~\ref{eq:snipLayernorm}, and $C$ is selected by original SNIP without normalization. (b) Error rate statistic of ImageNet classification in different spaces. We randomly sampled 500 models in each space to inspect their error distributions under redundant one-shot training. Cumulative ratio represents the percentage of sampled models with error below a certain value.}\label{fg:search_space}
\vskip -0.1in
\end{figure}

\paragraph{Effect of Normalized Proxy} We further analyze the importance of normalized isomer proxy in our twp-step selector $\mathcal{S}$. In the first step, we preserve a unique representative from each group of Transformer isomers using a normalized SNIP score $\mathcal{P}^\textup{\tiny norm}_\textup{\tiny SNIP}$, which forms a refined search space $B$ as in Fig.~\ref{fg:search_space:a}. As a comparison, we apply the original SNIP without layer normalization in first step to obtain the search space $C$. We follow the criteria of RegNet~\cite{radosavovic2020designing} to characterize the qualities of refined search spaces. We randomly sample 500 architectures from the entire search space $A$ to analyze performance distribution. The corresponding isomers in $B$ and $C$ are picked to keep the constant parameters for fair comparison. As shown in Fig.~\ref{fg:search_space:b}, the curves imply that the quality of $B$ covers $A$ and $C$ and the models in $C$ perform worst.

\begin{table*}[t]
\caption{Transfer learning on different downstream tasks with ImageNet pre-training.}\label{tab:transfer}
\vskip 0.1in
\begin{center}
\begin{sc}
\begin{tabular*}{0.96\linewidth}{l|@{\extracolsep{\fill}}c|cccccc}
\toprule
Model              & \#Params & CIFAR-10 & CIFAR-100 & Flowers & Cars & Pets & iNat-19 \\ \midrule
EfficientNet-B5    & 30M      & 98.7     & 91.1      & 98.5    & -    & -    & -       \\ Grafit ResNet-50   & 25M      & -        & -         & 98.2    & 92.5 & -    & 75.9    \\
\midrule
ViT-B/16           & 86M      & 98.1     & 87.1      & 89.5    & -    & 93.8 & -       \\
ViT-L/16           & 307M     & 97.9     & 86.4      & 89.7    & -    & -    & -       \\
DeiT-B             & 86M      & 99.1     & 90.8      & 98.4    & 92.1 & -    & 77.7    \\
ViTAE-S            & 24M      & 98.8     & 90.8      & 97.8    & 91.4 & 94.2 & 76.0    \\
DearKD-S           & 22M      & 98.4     & 89.3      & 97.4    & 91.3 & -    & -       \\ \midrule
\textbf{\name{}-S} & \textbf{23M}      & \textbf{99.1}     & \textbf{91.2}      & \textbf{97.6}    & \textbf{92.2} & \textbf{94.9} & \textbf{76.4}    \\ \bottomrule
\end{tabular*}
\end{sc}
\end{center}
\vskip -0.1in
\end{table*}

\begin{table}[tb]
\caption{Comparative results of CNN architectures on ImageNet. The paradigm of \cnnname{} is applied to BigNAS search space.}\label{tab:sotacnn}
\vskip 0.1in
\begin{center}
\begin{sc}
\begin{tabular}{clcc}
\toprule
Group   & Model               & \makecell{Top-1\\ (\%)}       & \makecell{FLOPs\\ (M)}       \\ \midrule
\multirow{2}*{\makecell{200M}}    & BigNAS-S            & 76.5          & 242    \\
& \textbf{\cnnname{}-S $_{CNN}$}          & \textbf{77.4}           & \textbf{237}    \\
 \midrule
\multirow{2}*{\makecell{400M}}    &  BigNAS-M       & 78.9      & 418   \\
& \textbf{\cnnname{}-M $_{CNN}$}  & \textbf{79.9}       & \textbf{413}   \\
 \midrule
\multirow{2}*{\makecell{600M}}    &  BigNAS-L       & 79.5      & 586   \\
& \textbf{\cnnname{}-L $_{CNN}$}  & \textbf{80.0}        & \textbf{528}   \\
 \midrule
\multirow{2}*{\makecell{1000M}}    &  BigNAS-XL       & 80.9      & 1040   \\
& \textbf{\cnnname{}-XL $_{CNN}$}  & \textbf{81.4}         & \textbf{986}   \\
 \midrule
\end{tabular}
\end{sc}
\end{center}
\vskip -0.2in
\end{table}


\paragraph{Effect of Performance Balancing} During one-shot training, the optimizer will update the learnable weights of a subnet chosen from the search space at each iteration. A basic sampling method is to uniformly choose candidate architectures from the preferred space $\widetilde{\mathcal{A}}$ with equal probability. However, this can lead to a suboptimal optimization of certain weights as the variable building factors are trained to different frequencies. As shown in Fig.~\ref{fg:balancing}, uniform sampling leads to poor performance of small models due to the uncommon selection of variable factors. As a contrast, we also experimented with the resampling method employed in redundant one-shot training, i.e., randomly construct new architectures  from the decomposed search space using variable factors as the minimal component. This helps to improve small models, but the performances of most subnets deteriorate due to the growing number of subnets for optimization. The balanced sampling ensures that the most critical embedding dimension and depth are fairly updated, thus achieving consistently superior performance.

\subsection{Transfer Learning Results} 

As the power of generalization is important for deep learning models, we further evaluate \name{} on various downstream tasks to measure its transfer learning capability. Following convention, the compared models are pre-trained on ImageNet and then fine-tuned on the target datasets. As shown in Tab.~\ref{tab:transfer}, we present results on CIFAR-10/100~\cite{cifar}, Flowers-102~\cite{flowers}, Stanford Cars~\cite{cars}, Oxford-IIIT Pets~\cite{pets}, and iNaturalist 2019~\cite{iNaturalist}. Our small model is able to outperform ViT and DeiT on multiple datasets with several-fold fewer parameters. \name{}-S is mostly superior to similar-scale ViTAE-S~\cite{vitae} and DearKD-S~\cite{dearkd} and is on par with convnet models.

\subsection{CNN Results}

Although the main focus of \name{} is on Vision Transformer, the concept of preferred learning for one-shot NAS is general and can be applied to various architectures. Here we demonstrate the adaptation on CNN architectures to prove its versatility and extensibility. We choose BigNAS~\cite{bignas} as the experimental basis, which is a popular single-stage NAS framework for CNN architectures. The paradigm of \name{} can be easily migrated to BigNAS, with the main effort being replacing search space of transformer blocks with convolutional layers. The experimented search space is detailed in Tab.~\ref{tab:BigNASSearchSpace}. Since isomers rarely appear in CNN architectures due to the down-sampling nature, the normalization step of Eq.~\ref{eq:snipLayernorm} can be skipped. The benchmark results are shown in Tab.~\ref{tab:sotacnn}.
Our \name{} achieves from $0.5\%$ to $1.0\%$ better accuracy for different model sizes than BigNAS in terms of similar FLOPs.

\FloatBarrier

\section{Conclusion}\label{conclusion}

In this paper, we proposed \name{}, a one-shot neural architecture search method that reduces the training space by identifying promising isomers and further pruning the training space through sparsification. By doing so, \name{} is able to conduct one-shot learning on a more focused set of architectures, leading to higher accuracy and is search-free after training by the zero-cost search in advance. Our experiments demonstrate that \name{} is able to produce highly accurate models with significantly reduced search times compared to other NAS methods. We believe that \name{} represents a promising approach to improving the efficiency and effectiveness of NAS, and we look forward to further exploring its potential in future research.




\bibliography{ref}
\bibliographystyle{icml2023}


\clearpage
\appendix
\onecolumn

\section{Statistical Analysis for SNIP with Weight Sharing Strategy in Transformer}
\label{appendix:snipAnalysis}
We analyze the subnets after the conventional one-shot training since it fairly optimizes the subnets in the whole search space and keeps the consistence of the performance with the model quality. We random sampling 2000 subnets with different model sizes inheriting weights form supernet and find that the SNIP score has 0.86 Kendall correlation with the performance, a littler higher than the 0.85 correlation of the model size and performance. To further leverage the reason of the bad performance of SNIP on the whole search space, we devote ourselves to analyzing the performance of SNIP on the subnets with equal model size.

\begin{table*}[h]
\caption{The statistics of the analyzed subnets in Case 1 and Case 2. We group the subnets by their size and make statistics of the number of subnets and the accuracy span from the worst subnet to the best subnet in the group.}\label{tab:snipAnalysisSample}
\vskip 0.1in
\begin{center}
\renewcommand{\arraystretch}{1.2}
\begin{tabular}{|c|l|ccccccccccc|}
\hline
\multirow{3}*{\text{\bf Case 1}} & Group ID & 1 & 2 & 3 & 4 & 5 & 6 & 7 & 8 & 9 & 10 & 11 \\
\cline{2-13}
& Model Num. & 12 & 66 & 220 & 495 & 792 & 924 & 792 & 495 & 220 & 66 & 12 \\
\cline{2-13}
& Acc. Span & 0.334 & 0.542 & 0.730 & 0.842 & 1.008 & 0.958 & 0.944 & 0.820 & 0.762 & 0.378 & 0.336 \\
\hline
\multirow{3}*{\text{\bf Case 2}} & Group ID & 1 & 2 & 3 & 4 & 5 & 6 & 7 & 8 & 9 & 10 & 11 \\
\cline{2-13}
& Model Num. & 12 & 66 & 220 & 495 & 792 & 924 & 792 & 495 & 220 & 66 & 12 \\
\cline{2-13}
& Acc. Span & 0.280 & 0.322 & 0.332 & 0.372 & 0.394 & 0.382 & 0.386 & 0.372 & 0.296 & 0.240 & 0.160 \\
\hline
\end{tabular}
\end{center}
\vskip -0.1in
\end{table*}

We focus our analysis on two cases:
\begin{enumerate}[label=(Case \arabic*),wide,labelindent=0pt]
\item We fix the embedding dimension as 192, depth as 12, number of heads in the multi-head attention block as 4 in each layer and consider all the subnets with 3.5 or 4 MLP ratio. We group the subnets as the Tab.~\ref{tab:snipAnalysisSample} shows where the group ID is equal to the number of layers with 4 MLP ratio.
\item We fix the embedding dimension as 192, depth as 12, MLP ratio as 4 in each layer and consider all the subnets with 3 or 4 heads in multi-head attention blocks. We group the subnets as the Tab.~\ref{tab:snipAnalysisSample} shows where the group ID is equal to the number of layers with 4 heads in multi-head attention blocks.
\end{enumerate}
As the Tab.~\ref{tab:snipAnalysisSample} shows, it is worthwhile to find a great distributions of the numbers of heads and especially of the MLP ratios since of the obvious different performance in the subnets with equal model sizes. Besides, that reduces the search space intensively while keeping the same range of model sizes as the whole search space. However, we find the SNIP \proxy{} fails in doing such things as following shows.

The mean Best Rank (mBR)~\cite{chen2021bench} is proposed for measuring the performance of the \zerocost{} \proxy{} in different sets of networks and emphasizes the actual quality of the best network selected by the \proxy{} which is very suitable for our situations. It can formalized as
$$
mBR = \frac{\sum_g (r_g - 1) }{\sum_g (|g| - 1)},
$$
where $g$ denotes the groups in our situations, $r_g$ is the actual rank in accuracy of the subnet selected by the \proxy{} as showed in Eq.~\ref{eq:zeroCost} and $|g|$ is the number of subnets in this group. If the \proxy{} always selects the best network, $mBR = 0$ but $mBR = 1$ for the worst network.

\begin{table*}[h]
\caption{The mBR of SNIP in Case 1 and Case 2.}\label{tab:snipAnalysismBR}
\vskip 0.1in
\begin{center}
\renewcommand{\arraystretch}{1.2}
\begin{tabular}{c|l|c}
\toprule
\multicolumn{2}{c|}{} & mBR \\
\midrule
\multirow{2}*{\text{\bf Case 1}} & SNIP & 0.986 \\
\cline{2-3}
& SNIP with layer norm. & 0.099 \\
\hline
\multirow{2}*{\text{\bf Case 2}} & SNIP & 0.825 \\
\cline{2-3}
& SNIP with layer norm. & 0.180 \\
\bottomrule
\end{tabular}
\end{center}
\vskip -0.1in
\end{table*}


As the Tab.~\ref{tab:snipAnalysismBR} shows, the SNIP selects almost the worst subnets in both Case 1 and Case 2. Then we compare the SNIP score in each layer of supernet. We find that there are higher SNIP scores in the lower layers since of the slight higher gradients. That leads SNIP to select the subnets with wider lower layers under the weight and gradient sharing strategy with supernet. In view of this, we propose the SNIP \proxy{} with layer normalization as the Eq.~\ref{eq:snipLayernorm} shows. The small mBRs of the SNIP \proxy{} with layer normalization in Tab.~\ref{tab:snipAnalysismBR} verify its effectiveness.

\section{The Greedy Algorithm}
\label{appendix:greedy}
The isomer architectures in $\mathcal{A}_g \in G$, where $G$ is the segmentation of the Transformer search space $\mathcal{A}$, have the same embedding dimensions, depth, total number of heads and total MLP ratios in all layers since of the same structures in layers of Transformer. Hence, we only need to consider how to allocate heads and MLP ratios to each layer to approximate the architecture with max $\mathcal{P}^\textup{\tiny norm}_\textup{\tiny SNIP}$ in each $\mathcal{A}_g$. Here, we only show the allocation for heads in Algorithm~\ref{alg:greedy} and the allocation of MLP ratio is similar. It is obvious that when there are only two choices of the number of heads and MLP ratio in each layer, the architecture induced by the greedy algorithm is exact the optimized one.

\begin{algorithm}[h]
   \caption{Greedy allocation of heads for isomer architectures}
   \label{alg:greedy}
\begin{algorithmic}
   \STATE {\bfseries Input:} The layer numbers $n$, total number $h$ of heads, head number lower bound $d$ and head number upper bound $u$.
   \STATE {\bfseries Output:} Head allocation in each layer $alc$.
   \STATE Initialize $alc$ as $alc[l] = d$ for each layer $l$.
   \STATE Calculate the number of left heads to allocate as $left = h - n*d$.
   \FOR{$i=1$ {\bfseries to} $left$}
        \STATE If $alc[l] < u$, calculate the SNIP score $\mathcal{P}[l]$ of the $(alc[l] + 1)$th head in each layer $l$ with layer normalization via Eq.~\ref{eq:snipLayernorm}, else $\mathcal{P}[l] = 0$.
        \STATE $cur = {\arg \max}_l \mathcal{P}[l]$
        \STATE $alc[cur] = alc[cur] + 1$
   \ENDFOR
   \STATE {\bfseries Return: $alc$}
\end{algorithmic}
\end{algorithm}

\section{Regularization \& Data Augmentation}\label{appendix:aug}

The learning capacity of three supernets are significantly different, with parameters ranging from minimum 5.4M to maximum 76M. Therefore, we appropriately decrease regularization and data augmentation for Supernet-Tiny and accordingly increase the magnitudes for Supernet-Base to avoid overfitting or underfitting. The detailed hyper-parameter settings are presented in Tab.~\ref{tab:hyper}. The value terminology primarily follows timm~\cite{timm}.

\begin{table}[h]
\caption{Hyper-parameters of training regularization and data augmentation.}\label{tab:hyper}
\vskip 0.1in
\centering
\renewcommand{\arraystretch}{1.2}
\begin{tabular}{lccc}
\toprule
Techniques        & Supernet-Tiny        & Supernet-Small    & Supernet-Base        \\ \midrule
Weight decay      & 0.02                 & 0.05     & 0.05        \\
Label smoothing   & 0.1                  & 0.1      & 0.1         \\
Stoch. Depth      & \cmark               & \cmark   & \cmark      \\
Repeated Aug      & \cmark               & \cmark   & \cmark      \\
\midrule
Mix switch prob       & \xmark     & 0.5      & 0.5         \\
Mixup alpha       & 0                  & 0.8      & 0.8         \\
Mixup mode        & \xmark                & elem     & elem        \\
Cutmix alpha      & 0                    & 1        & 1           \\
Rand Augment      & m9-n2-mstd0.5-inc1   & \xmark   & m10-n3-mstd0.5-inc1      \\
AutoAug           & \xmark               & v0r-mstd0.5      & \xmark      \\
Erasing prob      & 0.25                 & 0.25     & 0.25        \\
Erasing count     & 1                    & 1        & 2           \\
\bottomrule
\end{tabular}
\vskip -0.1in
\end{table}


\end{document}